\documentclass[letterpaper,twocolumn,10pt]{article}
\usepackage{usenix,epsfig,endnotes}
\usepackage{authblk}
\usepackage{graphicx}
\graphicspath{ {Figures/} }
\usepackage{multirow}
\usepackage{dblfloatfix} 
\usepackage{float} 
\usepackage{fixltx2e}
\usepackage{url}
\usepackage[pdfborder={0 0 0}]{hyperref}
\usepackage[caption=false]{subfig}
\usepackage{titling}
\usepackage{cite}
\usepackage{xspace}
\usepackage{mathtools}
\usepackage{booktabs}

\newcommand{\mycaption}[2]{\caption{\textbf{#1}. \textit{#2}}}
\newcommand{\sref}[1]{\S\ref{#1}}
\newcommand{\vheading}[1]{\vspace{0.05in}\noindent\textbf{#1}}
\newcommand{\fsync}{\texttt{fsync()}\xspace}

\newcommand{\eg}{\textit{e.g.,}\xspace}
\newcommand{\etc}{\textit{etc.}\xspace}
\newcommand*{\affaddr}[1]{#1} 
\newcommand*{\affmark}[1][*]{\textsuperscript{#1}}

\newcommand{\myx}{$\times$}
\newcommand{\vtt}[1]{\texttt{#1}}

\newcommand{\ra}[1]{\renewcommand{\arraystretch}{#1}}

\begin{document}

\date{}


\title{\Large \bf Analyzing IO Amplification in Linux File Systems
\vspace{0in}
}

\author{
{\rm Jayashree Mohan\affmark[1] }
\enspace 
{\rm Rohan Kadekodi\affmark[2] }
\enspace 
{\rm Vijay Chidambaram\affmark[1]}
\vspace{0em}  \\
\affaddr{\affmark[1]Department of Computer Science, University of Texas at Austin}\\
\affaddr{\affmark[2]Department of Computer Science, University of Wisconsin Madison}
\vspace{0in}
}
\maketitle


\subsection*{Abstract}

We present the first systematic analysis of read, write, and space
amplification in Linux file systems. While many researchers are
tackling write amplification in key-value stores, IO amplification in
file systems has been largely unexplored. We analyze data and
metadata operations on five widely-used Linux file systems: ext2, ext4,
XFS, btrfs, and F2FS. We find that data operations result in
significant write amplification (2--32\myx) and that metadata
operations have a large IO cost. For example, a single rename requires
648 KB write IO in btrfs. We also find that small random reads result
in read amplification of 2--13\myx. Based on these observations, we
present the CReWS conjecture about the relationship between IO
amplification, consistency, and storage space utilization. We hope
this paper spurs people to design future file systems with less IO
amplification, especially for non-volatile memory technologies.

\section{Introduction}
\label{sec-intro}

File systems were developed to enable users to easily and efficiently
store and retrieve data. Early file systems such as the Unix Fast File
System~\cite{mckusick1984fast} and ext2~\cite{mathur2007new} were
simple file systems. To enable fast recovery from crashes,
crash-consistency techniques such as
journaling~\cite{hagmann1987reimplementing} and
copy-on-write~\cite{HitzEtAl94-WAFL} were incorporated into file
systems, resulting in file systems such as ext4~\cite{Mathur+07-Ext4}
and xfs~\cite{sweeney1996scalability}. Modern file systems such as
btrfs~\cite{rodeh2013btrfs} include features such as snapshots and
checksums for data, making the file system even more complex.

While the new features and strong crash-consistency guarantees have
enabled wider adoption of Linux file systems, it has resulted in the
loss of a crucial aspect: efficiency. File systems now maintain a
large number of data structures on storage, and both data and metadata
paths are complex and involve updating several blocks on storage. In
this paper, we ask the question: what is the IO cost of various Linux
file-system data and metadata operations? What is the IO amplification
of various operations on Linux file systems? While this question is
receiving wide attention in the world of key-value
stores~\cite{bender2007cache, marmol2015nvmkv, wu2015lsm,
  sears2012blsm, shetty2013building, lu2016wisckey} and
databases~\cite{tokudb}, this has been largely ignored in file
systems. File systems have traditionally optimized for latency and
overall throughput~\cite{Chidambaram+12-NoFS, ChidambaramEtAl13-OptFS,
  PillaiEtAl17-CCFS, jannen2015betrfs}, and not on IO or space
amplification.

We present the first systematic analysis of read, write, and space
amplification in Linux file systems. Read amplification indicates the
ratio of total read IO to user data respectively. For example, if the
user wanted to read 4 KB, and the file system read 24 KB off storage
to satisfy that request, the read amplification is 6\myx. Write
amplification is defined similarly. Space amplification measures how
efficiently the file system stores data: if the user writes 4 KB, and
the file system consumes 40 KB on storage (including data and
metadata), the space amplification is 10\myx.

We analyze five widely-used Linux file systems that occupy different
points in the design space: ext2 (no crash consistency guarantees),
ext4 (metadata journaling), XFS (metadata journaling), F2FS
(log-structured file system), and btrfs (copy-on-write file
system). We analyze the write IO and read IO resulting from various
metadata operations, and the IO amplification arising from data
operations. We also analyze these measures for two macro-benchmarks:
compiling the Linux kernel, and Filebench
varmail~\cite{wilson2008new}. We break down write IO cost by IO that
was performed synchronously (during \fsync) and IO that was performed
during delayed background checkpointing.

We find several interesting results. For data operations such as
overwriting a file or appending to a file, there was significant write
amplification (2--32\myx). Small random reads resulted in a read
amplification of 2--8\myx, even with a warm cache. Metadata operations
such as directory creation or file rename result in significant
storage IO: for example, a single file rename required 12--648 KB to
be written to storage. Even though ext4 and xfs both implement
metadata journaling, we find XFS significantly more efficient for file
updates. Similarly, though F2FS and btrfs are implemented based on the
log-structured approach (copy-on-write is a dual of the log-structured
approach), we find F2FS to be significantly more efficient across all
workloads. In fact, in all our experiments, btrfs was an outlier,
producing the highest read, write, and space amplification. While this
may partly arise from the new features of btrfs (that other file
systems do not provide), the copy-on-write nature of btrfs is also
part of the reason.

We find that IO amplification arises due to three main factors: the
block-based interface, the crash-consistency mechanisms of file
systems, and the different data structures maintained by storage to
support features such as snapshots. Based on these observations, we
introduce the \emph{CReWS} conjecture. The CReWS conjecture states
that for a general-purpose file system on a shared storage device, it
is impossible to provide strong crash-consistency guarantees while
also minimizing read, write, and space amplification. We discuss
different designs of file systems, and show that for a general-purpose
file system (used by many applications), minimizing write
amplification leads to space amplification. We hope the CReWS
conjecture helps guide future file-system designers.

With the advent of non-volatile memory technologies such as Phase
Change Memory~\cite{raoux2008phase} that have limited write cycles,
file-system designers can no longer ignore IO amplification. Such
technologies offer the byte-based interface, which can greatly help to
reduce IO amplification. Data structures can be updated byte-by-byte
if required, and the critical metadata operations can be redesigned to
have low IO footprint. We hope this paper indicates the current state
of IO amplification in Linux file systems, and provides a useful guide
for the designers of future file systems.

\section{Analyzing Linux File Systems}

We now analyze five Linux file systems which represent a variety of
file-system designs. First, we present our methodology
(\sref{sec-meth}) and a brief description of the design of each file
system (\sref{sec-fs}). We then describe our analysis of common
file-system operations based on three aspects: read IO, write IO, and
space consumed (\S\ref{sec-analysis}).

\subsection{Methodology}
\label{sec-meth}

We use \vtt{blktrace}~\cite{blktrace}, \vtt{dstat}~\cite{dstat}, and
\vtt{iostat}~\cite{iostat} to monitor the block IO trace of different
file-system operations such as \vtt{rename()} on five different Linux
file systems. These tools allow us to accurately identify the
following three metrics.

\vheading{Write Amplification}. The ratio of total storage write IO to
the user data. For example, if the user wrote 4 KB, and that
resulted in the file system writing 8 KB to storage, the write
amplification is 2. For operations such as file renames, where there
is no user data, we simply report the total write IO. Write IO and
write amplification both should be minimized.

\vheading{Read Amplification}. Similar to write amplification, this is
the ratio of total storage read IO to user-requested data. For
example, if the user wants to read 4 KB, and the file system reads 12
KB off the storage to serve the read request, the read amplification is
3. We report the total read IO for metadata operations such as file
creation. Read amplification should also be minimized.

\vheading{Space Amplification}. The ratio of bytes consumed on storage
to bytes stored by the user. For example, the user wants to store 4
KB. If the file system has to consume 20 KB on storage to store 4 KB
of user data, the space amplification is 5. Space amplification is
a measure of how efficiently the file system is using storage, and thus
should be minimized. We calculate space amplification based on the
unique disk locations written to the storage, during the workloads.

Note that if the user stores one byte of data, the write and space
amplification is trivially 4096 since the file system performs IO in
4096 block-sized units. We assume that a careful application will
perform read and write in multiples of the block size. We also use
\vtt{noatime} in mounting the file systems we study. Thus, our results
represent amplification that will be observed even for careful
real-world applications.

\subsection{File Systems Analyzed}
\label{sec-fs}

We analyze five different Linux file systems. Each of these file
systems is (or was in the recent past) used widely, and represents a
different point in the file-system design spectrum.

\vheading{ext2}. The ext2 file system~\cite{CardEtAl94-Ext2} is a
simple file system based on the Unix Fast File
System~\cite{mckusick1984fast}. ext2 does not include machinery for
providing crash consistency, instead opting to fix the file system
with fsck after reboot. ext2 writes data in place, and stores file
metadata in inodes. ext2 uses direct and indirect blocks to find data
blocks.

\vheading{ext4}. ext4~\cite{mathur2007new} builds on the ext2
codebase, but uses journaling~\cite{hagmann1987reimplementing} to
provide strong crash-consistency guarantees. All metadata is first
written to the journal before being checkpointed (written in-place) to
the file system. ext4 uses extents to keep track of allocated blocks.

\vheading{XFS}. The XFS~\cite{sweeney1996scalability} file system also
uses journaling to provide crash consistency. However, XFS implements
journaling differently from ext4. XFS was designed to have high
scalability and parallelism. XFS manages the allocated inodes through
the inode B+ tree, while the free space information is managed by B+
trees. The inodes keep track of their own allocated extents.

\vheading{F2FS}. F2FS~\cite{lee2015f2fs} is a log-structured file
system designed specifically for solid state drives. Similar to the
original LFS~\cite{rosenblum1992design}, F2FS writes all updates to
storage sequentially. The logs in F2FS are composed of multiple segements, 
with the segment utilization monitored using Segment Information Table (SIT). 
Additionally, to avoid the wandering tree problem~\cite{bityutskiy2005jffs3}, F2FS assigns a node ID
 to the metadata structures like inodes, direct and indirect blocks. 
The mapping between node ID and the actual blockaddress is maintained in a Node Address Table (NAT),
which has to be referred to read data off storage, resulting in some overhead. 
Though data is written sequentially to the logs, NAT and SIT updates are first journaled and then written out in place.

\vheading{btrfs}. btrfs~\cite{rodeh2013btrfs} is a copy-on-write file
system based on B+ trees. The entire file system is composed of
different B+ trees (\eg file-system tree, extent tree, checksum tree,
\etc), all emerging from a single tree called as the tree of tree
roots. All the metadata of Btrfs is located in these trees. The
file-system tree stores the information about all the inodes, while
the extent tree holds the metadata related to each allocated
extent. Btrfs uses copy-on-write logging, in which any modification to
a B+ tree leaf/node is preceded by copying of the entire leaf/node to
the log tree.

\subsection{Analysis}
\label{sec-analysis}

We measure the read IO, write IO, and space consumed by different
file-system operations. 

\subsubsection{Data Operations}
First, we focus on data operations: file read,
file overwrite, and file append. For such operations, it is easy to
calculate write amplification, since the workload involves a fixed
amount of user data. The results are presented in
Table~\ref{tbl-data}.

\begin{table}[!t]
  \small
  \centering
    \ra{1.3}
  \begin{tabular}{@{}lrrrrr@{}}
    \toprule[1.2pt]
    Measure & ext2 & ext4 & xfs & f2fs & btrfs \\
    \midrule
    \emph{File Overwrite}\\
    Write Amplification & 2.00 & 4.00 & 2.00 & 2.66 & 32.65 \\
    Space Amplification & 1.00 & 4.00 & 2.00 & 2.66 & 31.17 \\
    \midrule
    \emph{File Append}\\
    Write Amplification & 3.00 & 6.00 & 2.01 & 2.66 & 30.85 \\
    Space Amplification & 1.00 & 6.00 & 2.00 & 2.66 & 29.77 \\
    \midrule
    \emph{File Read (cold cache)}\\
    Read Amplification & 6.00 & 6.00 & 8.00 & 9.00 & 13.00 \\
    \emph{File Read (warm cache)}\\
    Read Amplification & 2.00 & 2.00 & 5.00 & 3.00 & 8.00 \\
    \bottomrule[1.2pt]
  \end{tabular}
  \mycaption{Amplification for Data Operations}{The table shows the
    read, write, and space amplification incurred by different file
    systems when reading and writing files.}
  \label{tbl-data}
\end{table}

\vheading{File Overwrite}. The workload randomly seeks to a
4KB-aligned location in a 100 MB file, does a 4 KB write (overwriting
file data), then calls \fsync to make the data durable. The workload
does 10 MB of such writes. From Table~\ref{tbl-data}, we observe that
ext2 has the lowest write and space amplification, primarily due to
the fact that it has no extra machinery for crash consistency; hence
the overwrites are simply performed in-place. The 2\myx write
amplification arises from writing both the data block and the inode
(to reflect modified time). XFS has a similar low write amplification,
but higher space amplification since the metadata is first written to
the journal. When compared to XFS, ext4 has higher write and space
amplification: this is because ext4 writes the superblock and other
information into its journal with every transaction; in other words,
XFS journaling is more efficient than ext4 journaling. Interestingly,
F2FS has an efficient implementation of the copy-on-write technique,
leading to low write and space amplification. The roll-forward
recovery mechanism of F2FS allows F2FS to write only the direct node
block and data on every \fsync, with other data checkpointed
infrequently~\cite{lee2015f2fs}. In contrast, btrfs has a complex
implementation of the copy-on-write technique (mostly due to a push to
provide more features such as snapshots and stronger data integrity)
that leads to extremely high space and write amplification. When btrfs is 
mounted with the default mount options that enable copy-on-write and checksumming
 of both data and metadata, we see 32\myx  write amplification as shown
 in Table~\ref{tbl-data}. However, if the checksumming of the user data is disabled,
 the write amplification drops to 28\myx, and when the copy-on-write feature is also disabled for user data (metadata is still copied on write), the write amplification for overwrites comes down to about 18.6\myx. An interesting take-away from this analysis is that even if you
pre-allocate all your files on these file systems, writes will still
lead to 2--30\myx write amplification.

\vheading{File Append}. Our next workload appends a 4 KB block to the
end of a file and calls \fsync. The workload does 10 MB of such
writes. The appended file is empty initially. Our analysis for the
file overwrite workload mostly holds for this workload as well; the
main difference is that more metadata (for block allocation) has to be
persisted, thus leading to more write and space amplification for ext2
and ext4 file systems. In F2FS and xfs, the block allocation
information is not persisted at the time of \fsync, leading to
behavior similar to file overwrites. Thus, on xfs and f2fs,
pre-allocating files does not provide a benefit in terms of write
amplification.  

\begin{figure}[!t]
    \centering
    \includegraphics[width=0.45\textwidth]{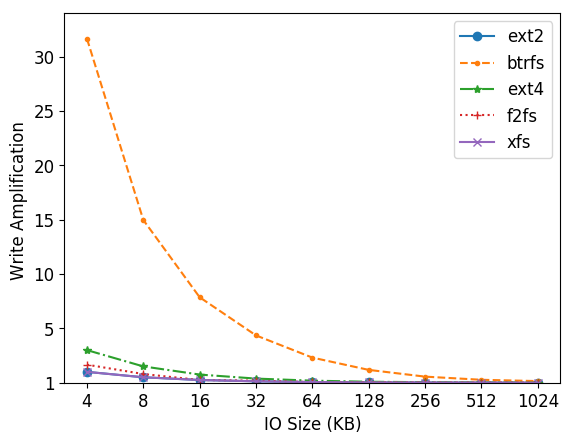}
    \vspace{-0.1in}
    \mycaption{Write Amplification for Various Write Sizes}{The figure
     shows the write amplification observed for writes of various
     sizes followed by a \fsync call.}
    \label{fig-wsize}
\end{figure}

We should note that write amplification is high in our workloads
because we do small writes followed by a \fsync. The \fsync call
forces file-system activity, such as committing metadata transactions,
which has a fixed cost regardless of the size of the write. As
Figure~\ref{fig-wsize} shows, as the size of the write increases, the
write amplification drops close to one. Applications which issue small
writes should take note of this effect: even if the underlying
hardware does not get benefit from big sequential writes (such as
SSDs), the file system itself benefits from larger writes. 

\vheading{File Reads}. The workload seeks to a random 4 KB aligned
block in a 10 MB and reads one block. In Table~\ref{tbl-data}, we make
a distinction between a cold-cache read, and a warm-cache read. On a
cold cache, the file read usually involves reading a lot of
file-system metadata: for example, the directory, the file inode, the
super block \etc. On subsequent reads (warm cache), reads to these
blocks will be served out of memory. The cold-cache read
amplification is quite high for all the file systems. Even in the case
of simple file systems such as ext2, reading a file requires reading
the inode. The inode read triggers a read-ahead of the inode table,
increasing the read amplification. Since the read path does not
include crash-consistency machinery, ext2 and ext4 have the same read
amplification. The high read amplification of xfs results from reading
the metadata B+ tree and readahead for file data. F2FS read
amplification arises from reading extra metadata structures such as
the NAT table and the SIT table~\cite{lee2015f2fs}. In btrfs, a
cold-cache file read involves reading the Tree of Tree roots, the
file-system and the checksum tree, leading to high read
amplification. On a warm cache, the read amplification of all file
systems greatly reduces, since global data structures are likely to be
cached in memory. Even in this scenario, there is 2--8\myx read
amplification for Linux file systems.

\begin{table}[!t]
  \small
  \centering
    \ra{1.3}
  \begin{tabular}{@{}lrrrrr@{}}
    \toprule[1.2pt]
    Measure & ext2 & ext4 & xfs & f2fs & btrfs \\
    \midrule
   \emph{File Create}\\
   Write Cost (KB) & 24 & 52 & 52 & 16 & 116 \\
   \multicolumn{1}{r}{\emph{fsync}} & 4 & 28 & 4 & 4 & 68 \\
      \multicolumn{1}{r}{\emph{checkpoint}} & 20 & 24 & 48 & 12 & 48 \\
    Read Cost (KB) & 24 & 24 & 32 & 36 & 40 \\
    Space Cost (KB) & 24 & 52 & 20 & 16 & 116 \\
    \midrule
   \emph{Directory Create}\\
    Write Cost (KB) & 28 & 64 & 80 & 20 & 132 \\
   \multicolumn{1}{r}{\emph{fsync}} & 4 & 36 & 4 & 8 & 68 \\
      \multicolumn{1}{r}{\emph{checkpoint}} & 24 & 28 & 76 & 12 & 64 \\
    Read Cost (KB) & 20 & 20 & 60 & 36 & 60 \\
    Space Cost (KB) & 28 & 64 & 54 & 20 & 132 \\
    \midrule
   \emph{File Rename}\\
    Write Cost (KB) & 12 & 32 & 16 & 20 & 648 \\
   \multicolumn{1}{r}{\emph{fsync}} & 4 & 20 & 4 & 8 & 392 \\
      \multicolumn{1}{r}{\emph{checkpoint}} & 8 & 12 & 12 & 12 & 256 \\
    Read Cost (KB) & 20 & 24 & 48 & 40 & 48 \\
    Space Cost (KB) & 12 & 32 & 16 & 20 & 392 \\
    \bottomrule[1.2pt]
  \end{tabular}
  \mycaption{IO Cost for Metadata Operations}{The table shows the
    read, write, and space IO costs incurred by different file
    systems for different metadata operations. The write cost is
    broken down into IO at the time of \fsync, and checkpointing IO
    performed later.}
  \label{tbl-meta}
\end{table}

\subsubsection{Metadata Operations}
We now analyze the read and write IO (and space consumed) by different
file-system operations. We present file create, directory create, and
file rename. We have experimentally verified that the behavior of
other metadata operations, such as file link, file deletion, and
directory deletion, are similar to our presented
results. Table~\ref{tbl-meta} presents the results. Overall, we find
that metadata operations are very expensive: even a simple file rename
results in the 12--648 KB being written to storage. On storage with
limited write cycles, a metadata-intensive workload may wear out the
storage quickly if any of these file systems are used.

In many file systems, there is a distinction between IO performed at
the time of the \fsync call, and IO performed later in the
background. The \fsync IO is performed in the critical path, and thus
contributes to user-perceived latency. However, both kinds of IO
ultimately contribute to write amplification. We show this breakdown
for the write cost in Table~\ref{tbl-meta}. 

\vheading{File Create}. The workload creates a new file in a
pre-existing directory of depth three (\eg \vtt{a/b/c}) and calls
\fsync on the parent directory to ensure the creation is persisted.
File creation requires allocating a new inode and updating a
directory, and thus requires 16--116 KB of write IO and 24--40 KB of
read IO in the various file systems. F2FS is the most efficient in
terms of write IO (but requires a lot of read IO). Overall, ext2 is
the most efficient in performing file creations. ext2, XFS, and F2FS
all strive to perform the minimum amount of IO in the \fsync critical
path. Due to metadata journaling, ext4 writes 28 KB in the critical
path. btrfs performs the worst, requiring 116 KB of write IO (68 KB in
the critical path) and 40 KB in checkpointing IO. The poor performance
of btrfs results from having to update a number of data structures,
including the file-system tree, the directory index, and
backreferences to create a file~\cite{rodeh2013btrfs}.

\vheading{Directory Create}. The workload creates a new directory in
an existing directory of depth four, and calls \fsync on the parent
directory. Directory creation follows a similar trend to file
creation. The main difference is the additional IO in creating the
directory itself. As before, btrfs experience the most write IO cost
and read IO cost for this workload. ext2 and F2FS are the most
efficient.

\vheading{File Rename}. The workload renames a file within the same
directory, and calls \fsync on the parent directory to ensure the
rename is persisted. Renaming a file requires updating two
directories. Performing rename atomically requires machinery such as
journaling or copy-on-write. ext2 is the most efficient, requiring
only 32 KB of IO overall. Renaming a file is a surprisingly complex
process in btrfs. Apart from linking and unlinking files, renames also
change the backreferences of the files involved. btrfs also logs the
inode of every file and directory (from the root to the parent
directory) involved in the operation. The root directory is persisted
twice, once for unlink, and once for the link. As a result, btrfs is
the least efficient, requiring 696 KB of IO to rename a single
file. Even if many of these inodes are cached, btrfs renames are
significantly less efficient than in other file systems.

\vheading{Macro-benchmark: Kernel Compilation}. To provide a more
complete picture of the IO amplification of file systems, we also
measure IO amplification for a macro-benchmark: uncompressing a Linux
kernel tarball, and compiling the kernel. The results are presented in
Table~\ref{tbl-macro}. The file systems perform 6.09--6.41 GB of write
IO and 0.23--0.27 GB of read IO. ext2 is the most efficient file
system, achieving the lowest write and space cost. Among file systems
providing crash-consistency guarantees, ext4 and XFS perform well,
achieving lower write and space cost than the copy-on-write file
systems of F2FS and btrfs. btrfs performs the most write IO, and uses
the most space on storage. The kernel compilation workload does not
result in lot of write amplification (or variation between file
systems), because the \fsync is not called often; thus each file
system is free to group together operations to reduce IO and space
cost. Even in this scenario, the higher write and space amplification
of btrfs is observed.

\vheading{Macro-benchmark: Filebench Varmail}. We ran the Varmail
benchmark from the Filebench benchmark suite~\cite{wilson2008new} with
the following parameters: 16 threads, total files 100K, mean file size
16 KB. Varmail simulates a mail server, and performs small writes
followed by \fsync on different files using multiple threads. In this
\fsync-heavy workload, we see that the effects of write, read, and
space amplification are clear. ext2 still performs the least IO and
uses the least storage space. btrfs performs 38\% more write IO than
ext2, and uses 39\% more space on storage. F2FS performs better than
btrfs, but has a high read cost (10\myx other file systems).

\begin{table}[!t]
  \small
  \centering
    \ra{1.3}
  \begin{tabular}{@{}lrrrrr@{}}
    \toprule[1.2pt]
    Measure & ext2 & ext4 & xfs & f2fs & btrfs \\
    \midrule
   \emph{Kernel Compilation}\\
    Write Cost (GB) & 6.09 & 6.19 & 6.21 & 6.38 & 6.41 \\
    Read Cost (GB) & 0.25 & 0.24 & 0.24 & 0.27 & 0.23 \\
    Space Cost (GB) & 5.94 & 6.03 & 5.96 & 6.2 & 6.25\\
    \midrule
   \emph{Filebench Varmail}\\
    Write Cost (GB) & 1.52 & 1.63 & 1.71 & 1.82 & 2.10 \\
    Read Cost (KB) & 116 & 96 & 116 & 1028 & 0 \\
    Space Cost (GB) & 1.45 & 1.57 & 1.50 & 1.77 & 2.02\\
    \bottomrule[1.2pt]
  \end{tabular}
  \mycaption{IO Cost for Macro-benchmarks}{The table shows the read,
    write, and space IO costs incurred by different file systems when
    compiling the Linux kernel 3.0 and when running the Varmail
    benchmark in the Filebench suite.}
  \label{tbl-macro}
\end{table}

\vheading{Discussion}. IO and space amplification arises in Linux file
systems due to using the block interface, from crash-consistency
techniques, and the need to maintain and update a large number of data
structures on storage. Comparison of XFS and ext4 shows that even when
the same crash-consistency technique (journaling) is used, the
implementation leads to a significant difference in IO
amplification. With byte-addressable non-volatile memory technologies
arriving on the horizon, using such block-oriented file systems will
be disastrous. We need to develop lean, efficient file systems where
operations such as file renames will result in a few bytes written to
storage, not tens to hundreds of kilobytes. 

\section{The CReWS Conjecture}
Inspired by the RUM conjecture~\cite{rum} from the world of key-value
stores, we propose a similar conjecture for file systems: the CReWS
conjecture\footnote{We spent some time trying to come up with
  something cool like RUM, but alas, this is the best we could do}.

\emph{The CReWS conjecture states that it is impossible for a
  general-purpose file system to provide strong crash (C)onsistency
  guarantees while simultaneously achieving low (R)ead amplification,
  (W)rite amplification, and (S)pace amplification.}

By a general-purpose file system we mean a file system used by
multiple applications on a shared storage device. If the file system
can be customized for a single application on a dedicated storage
device, we believe it is possible to achieve the other four properties
simultaneously.

For example, consider a file system designed specifically for an
append-only log such as Corfu~\cite{balakrishnan2012corfu} (without
the capability to delete blocks). The storage device is dedicated for
the append-only log. In this scenario, the file system can drop all
metadata and treat the device as a big sequential log; storage block 0
is block 0 of the append-only log, and so on. Since there is no
metadata, the file system is consistent at all times implicitly, and
there is low write, read, and space amplification. However, this only
works if the storage device is completely dedicated to one
application.

Note that we can extend our simple file-system to a case where there
are N applications. In this case, we would divide up the storage into
N units, and assign one unit to each application. For example, lets
say we divide up a 100 GB disk for 10 applications. Even if an
application only used one byte, the rest of its 10 GB is not available
to other applications; thus, this design leads to high space
amplification.

In general, if multiple applications want to share a single storage
device without space amplification, dynamic allocation is
required. Dynamic allocation necessitates metadata keeping track of
resources which are available; if file data can be dynamically
located, metadata such as the inode is required to keep track of the
data locations. The end result is a simple file system such as
ext2~\cite{CardEtAl94-Ext2} or NoFS~\cite{Chidambaram+12-NoFS}. While
such systems offer low read, write, and space amplification, they
compromise on consistency: ext2 does not offer any guarantees on a
crash, and a crash during a file rename on NoFS could result in the
file disappearing.

File systems that offer strong consistency guarantees such as ext4 and
btrfs incur significant write amplification and space amplification,
as we have shown in previous sections. Thus, to the best of our
knowledge, the CReWS conjecture is true.

\vheading{Implications}. The CReWs conjecture has useful implications
for the design of storage systems. If we seek to reduce write
amplification for a specific application such as a key-value store, it
is essential to sacrifice one of the above aspects. For example, by
specializing the file system to a single application, it is possible
to minimize the three amplification measures. For applications seeking
to minimize space amplification, the file system design might
sacrifice low read amplification or strong consistency guarantees. For
non-volatile memory file systems~\cite{xu2016nova, dulloor2014system},
given the limited write cycles of non-volatile
memory~\cite{zhou2009durable}, file systems should be designed to
trade space amplification for write amplification; given the high
density of non-volatile memory technologies~\cite{raoux2008phase,
  xue2011emerging, ho2009nonvolatile, strukov2008missing,
  chua2011resistance}, this should be acceptable.  Thus, given a goal,
the CReWS conjecture focuses our attention on possible avenues to
achieve it.

\section{Conclusion}
\label{sec-conc}

We analyze the read, write, and space amplification of five Linux file
systems. We find that all examined file systems have high write
amplification (2--32\myx) and read amplification (2--13\myx). File
systems that use crash-consistency techniques such as journaling and copy-on-write
also suffer from high space amplification (2--30\myx). Metadata
operations such as file renames have large IO cost, requiring 32--696
KB of IO for a single rename.

Based on our results, we present the CReWS conjecture: that a
general-purpose file system cannot simultaneously achieve low read,
write, and space amplification while providing strong consistency
guarantees. With the advent of byte-addressable non-volatile memory
technologies, we need to develop leaner file systems without
significant IO amplification: the CReWS conjecture will hopefully
guide the design of such file systems.

{\footnotesize \bibliographystyle{unsrt}
\bibliography{all}}

\end{document}